\iftrue 
\documentclass[final,1p,times]{elsarticle}
\else
\documentclass[final,1p,times,dvipdfmx]{elsarticle}
\fi
\journal{Magnetic Resonance Imaging}

\usepackage[colorlinks,bookmarksopen,bookmarksnumbered,citecolor=red,urlcolor=red,breaklinks=true]{hyperref}
\usepackage[colorlinks,bookmarksopen,bookmarksnumbered,citecolor=red,urlcolor=red,breaklinks=true]{hyperref}

\usepackage{amsmath}

\newtheorem{remark}{Remark}

\def\pp#1#2{\frac{\partial #1}{\partial #2}}
\def\bm#1{\boldsymbol{#1}}
\def\diff{\mathrm{d}}

\usepackage{siunitx}

\NewDocumentCommand\nprounddigits{m}{\sisetup{round-precision = #1}}
\def\npproductsign#1{} 
\def\reason#1{(\because~\text{#1})}

\def\vB{\bm{B}} 
\def\vx{\bm{x}}
\def\vs{\bm{s}}
\def\vz{\bm{z}}
\def\vV{\bm{V}}
\def\vr{\bm{r}}
\def\vP{\bm{P}}
\def\ve{\bm{e}}

\def\Oee{O(\epsilon^2)}

\def\tilde#1{\widetilde{#1}}

\begin{document}

\begin{frontmatter}
  \title{Designing gradient coils with the shape derivative and the closed B-spline curves}
  \author[NU]{Toru Takahashi\corref{cor}}
  \ead{toru.takahashi@mae.nagoya-u.ac.jp}
  \cortext[cor]{Corresponding author}
  \address[NU]{Department of Mechanical Systems Engineering, Nagoya University, Furo-cho, Chikusa-ku, Nagoya city, Aichi, 464-8603 Japan}
  
  \begin{abstract}
    This study proposes a versatile and efficient optimisation method for discrete coils that induce a magnetic field by their steady currents. The prime target is gradient coils for MRI (Magnetic Resonance Imaging). The derivative (gradient) of the $z$-component the magnetic field, which is calculated by the Biot--Savart's law, with respect to the $z$-coordinate in the Cartesian $xyz$ coordinate system is considered as the objective function. Then, the derivative of the objective function with respect to a change of coils in shape is formulated according to the concept of shape optimisation. The resulting shape derivative (as well as the Biot--Savart's law) is smoothly discretised with the closed B-spline curves. In this case, the control points (CPs) of the curves are naturally selected as the design variables. As a consequence, the shape derivative is discretised to the sensitivities of the objective function with respect to the CPs. Those sensitivities are available to solve the present shape-optimisation problem with a certain gradient-based nonlinear-programming solver. The numerical examples exhibit the mathematical reliability, computational efficiency, and engineering applicability of the proposed methodology based on the shape derivative/sensitivities and the closed B-spline curves.
  \end{abstract}
  
\end{frontmatter}

\section{Introduction}\label{s:intro}

In MRI (Magnetic Resonance Imaging), gradient coils fulfil an important role to generate a uniform gradient magnetic field over a domain of interest (DOI). As a consequence, a number of numerical methods to analyse and design gradient coils have been developed so far. The design methods are classified to (I) discrete-coils approach and (II) current-density approach \cite{turner1993,hidalgo2010,smith2016}.

The second approach (II) optimises the distribution of current in conductors in a prescribed spatial configuration and includes the target field approach \cite{turner1986target_field_approach}, the equivalent magnetic dipole method \cite{shen2022gradient_coil_design}, and so on. This approach is more elaborated and thus powerful than the first approach as mentioned below. As a drawback, it is computationally more expensive because a certain mesh-based method (typically, finite element method \cite{ryu2006IEEE} and boundary element method \cite{sanchez2011design}) is necessary to solve the Ampere's law (or its equivalence) in terms of the vector potential \cite{fernow2022principles}. Moreover, the current-density methods are an indirect approach, which optimises indirect parameters such as stream function \cite{shen2022gradient_coil_design}. Since the present study is categorised to the first approach, individual methods for the second approach are left to the review articles \cite{turner1993,hidalgo2010}.

The first approach (I) includes the designs of the Helmholtz pair coil \cite{fernow2022principles} (which is for generating a uniform field), the Maxwell pair coil \cite{hidalgo2010}, the Golay-type gradient coil \cite{golay1958}, and so on. As a development of Golay \cite{golay1958}, Rom\'eo et al. \cite{romeo1984} proposed to create a desired magnetic field with axially symmetric coils by considering the properties of the higher terms of the field expansion in terms of the spherical harmonics. These studies are combinatorial designs rather than mathematical optimisations of discrete coils.

Another subgroup in (I) relies various optimisation techniques to find optimised parameters regarding geometry of gradient coils. In this subgroup, it is commonly assumed that steady currents run through thin coils or wires. Thus, the (steady-state) magnetic field can be computed by the Biot--Savart's law. Hence, since the magnetic field is represented analytically (as in Eq.~(\ref{eq:B})), the derivative (gradient) of a given objective function, which is usually related to the magnetic field, can be calculated with respect to a certain parameter regarding the geometrical configuration of the target coil. This enables to apply a variety of gradient-based optimisation methods to the underlying optimisation problems in order to obtain a uniform gradient magnetic field of interest. Many investigations consider the parameters specific to a given configuration of coils \cite{wong1991,du1997studies,fisher1997design,juchem2010,du2012designCT,zhang2018optimisation,zhang2018spiral,xuan2021}.

On contrast to the above parametric optimisations focusing on a specific type of coils \cite{wong1991,juchem2010,du2012designCT,zhang2018optimisation,zhang2018spiral,xuan2021}, a more flexible and general-purpose methodology is to approximate a coil with small portions, which are called current/wire elements, and then find their best positions. For this regard, Brey et al. \cite{brey1996} applied the CGD method to optimise locations and/or currents of wire elements. The gradient necessary for the CGD method can be calculated by differentiating the magnetic field with respect to the (cylindrical) coordinates of the elements. Likewise, Lu et al. \cite{lu2004momentum-weighted} optimised the positions of wire elements with the momentum-weighted CGD algorithm to take care of the issues of the CGD methods. A possible drawback of optimising the positions of wire elements is to handle a vast number of design variables as the number of such elements increases. This inefficiency is conceivably a serious issue if a coil was complicated in shape because it could consist of many wire elements.

As a more efficient method, this study proposes to approximate a coil with a closed B-spline curve. It is reasonable to choose the control points (CPs) as the design variables. If the degree of B-spline functions is one, the resulting closed B-spline curve is piecewise-linear, which is equivalent to the existing wire elements \cite{brey1996,lu2004momentum-weighted}. Otherwise, when a CP is moved, the closed B-spline curve can deform smoothly and locally near the CP. This feature of B-spline functions, which can be generalised to NURBS functions, has been utilised to design surfaces (rather than lines) in shape optimisations using mesh-based PDE solvers such as the isogeometric analysis \cite{hughes2005}. With regard to the MRI, the surface shape of the pole piece in a permanent magnet MRI system was designed with using 2D B-spline functions by Ryu et al. \cite{ryu2006IEEE}, although this study is classified to (II).

When expressing a coil with a closed B-spline curve, it will be revealed that the shape sensitivity of a certain objective function, which is related to the quantities for the magnetic field, can be calculated with respect to each CP. To this end, the shape derivative, which is the Eulerian derivative \cite{sokolowski1991} of an objective function with respect to the translation in an arbitrary direction at any point on the coil, is derived from the Biot--Savart's law by considering the perturbation of the point. The derived shape sensitivities are integrated to a general-purpose nonlinear-programming solver to optimise the positions of the design variables, i.e. CPs. As far as the author knows, this kind of NURBS- or B-spline-based optimisation method has been unreported not only for designing gradient coils but also for the other coils/wires.

In what follows, Section~\ref{s:gradient} constructs the B-spline-based shape optimisation method to find a gradient coil such that $\pp{B_z}{z}$, i.e. the gradient in the $z$-direction of the $z$-component of the magnetic field $\vB$, is close to the desired gradient at given target points. Section~\ref{s:sd_grad} rigorously verifies the shape derivative (sensitivity), which is the core of the proposed shape optimisation method. Section~\ref{s:gradcoil} solves some shape optimisation problems for $z$-gradient coils with the proposed coil-design method. Section~\ref{s:conclusion} concludes this study.

\section{Coil design method}\label{s:gradient}

In this section, an optimisation problem will be considered to find a uniform gradient of the magnetic field $\vB$, which is the prime task in designing gradient coils. After the problem statement, the shape derivative of the objective function of interest will be derived theoretically and discretised with the B-spline functions. Finally, the reduction to a nonlinear programming will be described.

\subsection{Problem setting}

\begin{figure}[hbt]
  \centering
  \includegraphics[width=.4\textwidth]{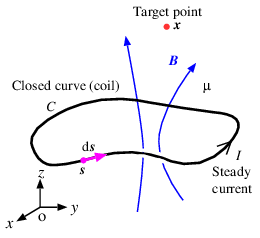}
  \caption{Problem setting for the optimisation problem in Section~\ref{s:gradient}.}
  \label{fig:problem}
\end{figure}

Suppose that a steady current $I$ flows through a coil $C$, which is an ideal of thin wire, in 3D (Figure~\ref{fig:problem}). Here, although $C$ may consist of multiple closed curves and they have different currents and shapes, a single closed curve will be simply considered in the following description. The current $I$ induces the magnetic field $\vB$ at any point $\vx$, which is assumed to be not on $C$ and will be called \textit{target point} hereafter. As well known, the field is governed by the following Biot--Savart's law \cite{feynman1965flp}:
\begin{align}
  \vB(\vx)=\frac{\mu I}{4\pi}\int_C\frac{\diff\vs\times(\vx-\vs)}{|\vx-\vs|^3},
  \label{eq:B}
\end{align}
where $\vs$ denotes a (source) point on $C$ and $\mu$ denotes the permeability of the underlying magnetic field. Here, it should be noted that, since $\bm{x}$ is not included in $C$, the vector $\vx-\vs$ never becomes zero and thus the integral is non-singular.  Then, let us make $B_z$, i.e. the $z$-component of $\vB$, have a target value, denoted by $g$, of the gradient in the $z$-direction at a target point $\vx$ in the Cartesian coordinates $\textrm{O}-xyz$. To this end, one may minimise an objective function
\begin{subequations}
  \begin{align}
    K(C)&:=\frac{1}{2}\left(\frac{\partial B_z(\bm{x})}{\partial z} - g \right)^2.
    \label{eq:J grad single}\\
    \intertext{More generally, it is easy to consider multiple target points, i.e. $\vx_1$, $\vx_2$, and so on, instead of the single point $\vx$. In this case, the objective function can be written as}
    K(C)&:=\frac{1}{2}\sum_i\left(\frac{\partial B_z(\bm{x}_i)}{\partial z} - g_i \right)^2,
    \label{eq:J grad general}
  \end{align}%
  \label{eq:J grad}%
\end{subequations}%
where $g_i$ denotes a desired target gradient at $\vx_i$. For brevity, the simpler expression in (\ref{eq:J grad single}) will be considered in the following theoretical description, while (\ref{eq:J grad general}) will be considered in the numerical examples in Section~\ref{s:gradcoil}.

\begin{remark}
  It is possible to handle any combinations of the component of $\vB$ and the direction of the gradient, e.g. $\pp{B_x}{y}$. However, this study considers the typical case of $\pp{B_z}{z}$ in order to simplify the formulation.
\end{remark}

\subsection{Shape derivative}

\def\vez{\hat{\vz}}

\begin{figure}[hbt]
  \centering
  \includegraphics[width=.4\textwidth]{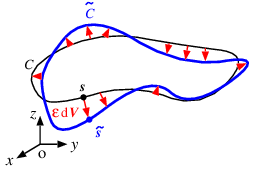}
  \caption{Perturbation of a closed curve $C$ by $\epsilon\vV(\vs)$ at $\vs$.}
  \label{fig:tilde C}
\end{figure}

One wants to know how $K$ in (\ref{eq:J grad}) changes when $C$ slightly changes in shape (Figure~\ref{fig:tilde C}). The change in $K$ is called the shape derivative or shape sensitivity of $K$ with respect to $C$. To formulate the shape derivative, one considers that any point $\vs$ on $C$ is perturbed to $\tilde{\vs}$ by an infinitesimal translate $\epsilon\vV(\vs)$, that is,
\begin{align}
  \tilde{\vs}:=\vs+\epsilon\vV(\vs),
  \label{eq:tilde s}
\end{align}
where $\epsilon$ denotes an infinitesimal number and $\vV=\vV(\bm{s})$ denotes the direction of the perturbation at $\vs$.

Correspondingly to $\tilde{\vs}$, the perturbed objective function $K(\tilde{C})$ is represented as
\begin{align}
  K(\tilde{C}) = K(C) + 2\left(\pp{B_z}{z}-g\right)\epsilon\frac{\mu I}{4\pi}\int_C\left[3(\diff\vs\times\vr)_z\left(\frac{V_z}{r^5}-\frac{5r_z}{r^7}\vV\cdot\vr\right)-\frac{3r_z(\diff\vV\times\vr-\diff\vs\times\vV)_z}{r^5}\right] + \Oee,
  \label{eq:J grad update}
\end{align}
where $\vr:=\vx-\vs$ and $r:=|\vr|$ are defined. The derivation of (\ref{eq:J grad update}) is described in \ref{s:derive_sd_grad}. Consequently, the shape sensitivity, denoted by $E=E(C,\vV)$, is defined and can be calculated as follows:
\begin{align}
  E(C,\vV)
  &:= \lim_{\epsilon\to 0}\frac{K(\tilde{C})-K(C)}{\epsilon}\nonumber\\
  &= \left(\pp{B_z}{z}-g\right)\frac{\mu I}{4\pi}\int_C\left[3(\diff\vs\times\vr)_z\left(\frac{V_z}{r^5}-\frac{5r_z}{r^7}\vV\cdot\vr\right)-\frac{3r_z(\diff\vV\times\vr-\diff\vs\times\vV)_z}{r^5}\right].
  \label{eq:J grad sd}
\end{align}

It should be noted that the differentiation of $B_z$ with respect to $z$ in (\ref{eq:J grad sd}) as well as (\ref{eq:J grad}) can be calculated as follows:
\begin{align*}
  \pp{B_z}{z}
  =\frac{\mu I}{4\pi}\int_C(\diff\vs\times\vr)_z\pp{}{z}\left(\frac{1}{r^3}\right)
  =\frac{\mu I}{4\pi}\int_C(\diff\vs\times\vr)_z\frac{-3r_z}{r^5},
\end{align*}
where it is used that $(\diff\vs\times\vr)_z$ does not depend on $z$.

\subsection{Discretisation with the B-spline functions}

It is necessary to discretise the Biot--Savart's law in (\ref{eq:B}) and the shape derivative $E$ in (\ref{eq:J grad sd}) to evaluate them numerically. To this end, this study approximates a coil with a closed B-spline curve.

\subsubsection{Expression of a coil with the closed B-spline curve}\label{s:closed B-spline}

\begin{figure}[hbt]
  \centering
  \includegraphics[width=.4\textwidth]{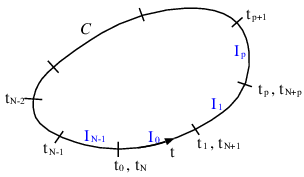}
  \caption{Knots $\{t_0,\ldots,t_N\}$ of a closed B-spline curve $C$ and knot intervals $\{I_0,\ldots,I_{N-1}\}$. The last $p$ knots, i.e. $t_{N+1}$, $\ldots$, $t_{N+p}$, are necessary to close the curve as explained in \ref{s:bspline}. The curve $C$ is drawn by increasing the parameter $t$ from $t_0$ to $t_N$ in (\ref{eq:s with R}).}
  \label{fig:C-discretise}
\end{figure}

One can express any point $\vs$ on $C$ as follows \cite{hoschek1993}:
\begin{align}
  \vs(t)=\sum_{n=0}^{N-1} R_n^p(t) \vP_n,
  \label{eq:s with R}
\end{align}
where $t$ denotes an arbitrary curve parameter along $C$, $\vP_0,\ldots,\vP_{N-1}$ denote the control points (CPs), and $R_0^p(t),\ldots,R^p_{N-1}(t)$ are the \textit{periodic B-spline functions} of degree $p$. These functions are associated with $n+p+1$ knots, i.e. $t_0,\ldots,t_{N+p}$  (Figure~\ref{fig:C-discretise}). For simplicity, the superscript $p$ of $R_n^p$ will be omitted hereafter. In addition, uniform knots such as
\begin{align*}
  t_k:=k \Delta\quad\text{for $k=0,\ldots,N+p$}
\end{align*}
will be assumed. Here, $\Delta:=\frac{1}{N}$ denotes the span between two adjacent knots. The details to construct the closed B-spline curve is described in \ref{s:bspline}.

For conciseness, the $k$-th interval on $C$ is defined by
\begin{align*}
  I_k : = [t_k,t_{k+1}]
\end{align*}
for $k=0,\ldots,N-1$; see Figure~\ref{fig:C-discretise}, again. Then, the contour integral over $C$ is split to $N$ intervals. The integral over an interval is evaluated by using $G$-points Gauss--Legendre quadrature method \cite{abramowitz1972}. See the details in \ref{s:integral}.

\subsubsection{Shape sensitivities of $K$ with respect to CPs}

Correspondingly to $\vs$ in (\ref{eq:s with R}), one can express $\tilde{\vs}$ in (\ref{eq:tilde s}) as 
\begin{align}
  \tilde{\vs}(t)=\sum_{n=0}^{N-1} R_n(t)\tilde{\vP}_n,
  \label{eq:tilde s with R}
\end{align}
which yields
\begin{align}
  \tilde{\vs}(t)-\vs(t)
  = \sum_{n=0}^{N-1} R_n(t)(\tilde{\vP}_n-\vP_n)
  = \sum_{n=0}^{N-1} R_n(t)\delta\vP_n.
\end{align}
Here, $\delta\vP_n:=\tilde{\vP}_n-\vP_n$ denotes the perturbation of $\vP_n$. This can be interpreted as the infinitesimal perturbation $\epsilon\vV(\vs)$ in (\ref{eq:tilde s}) \cite{takahashi2019ewco,takahashi2023IJNME,takahashi2024ewco_pre}, that is,
\begin{align}
  \epsilon\vV(\vs)=\sum_{n=0}^{N-1} R_n(t)\delta{\vP}_n.
  \label{eq:eV}
\end{align}
Consequently, the substitution of (\ref{eq:eV}) into (\ref{eq:J grad update}) leads to the following expression:
\begin{align}
  K(\tilde{C}) = K(C)+\sum_{n=0}^{N-1}\ve_n\cdot\delta\vP_n + \Oee,
  \label{eq:tilde J grad}
\end{align}
where
\begin{align}
  \ve_n
  &=\left(\pp{B_z(\vx)}{z}-g\right)\frac{\mu I}{4\pi}\int_C\left[
    \frac{3}{r^5} \begin{pmatrix}0 \\ 0 \\ (\dot{\vs}\times\vr)_z R_n \end{pmatrix}\
    -\frac{15(\dot{\vs}\times\vr)_z r_z R_n}{r^7}\vr
    -\frac{3r_z}{r^5}\begin{pmatrix} \dot{R}_n r_y + R_n \dot{s}_y \\ - \dot{R}_n r_x - R_n \dot{s}_x\\  0 \end{pmatrix}
    \right]\diff t\nonumber\\
  &=\left(\pp{B_z(\vx)}{z}-g\right)\frac{\mu I}{4\pi}\int_C \frac{3}{r^5}\left[
    \begin{pmatrix} - r_z(\dot{R}_n r_y + R_n \dot{s}_y) \\ r_z (\dot{R}_n r_x + R_n \dot{s}_x) \\  R_n(\dot{s}_x r_y-\dot{s}_y r_x) \end{pmatrix}
    -\frac{5(\dot{s}_x r_y - \dot{s}_y r_x)r_z R_n}{r^2}\vr
    \right]\diff t.
  \label{eq:en}
\end{align}
The three-dimensional vector $\ve_n$ represents the \textit{shape sensitivity} regarding the CP $\vP_n$. The derivation of (\ref{eq:tilde J grad}) is given in \ref{s:derive_sensitivity_grad}.

\subsection{Utilisation of a nonlinear programming solver}\label{s:nlopt}

The expression in (\ref{eq:tilde J grad}) indicates that the present optimisation problem is solved with a nonlinear programming solver that exploits the gradient of an objective function. In the present case of (\ref{eq:tilde J grad}), one can regard the $N$ CPs as the design variables to be sought and the shape sensitivities $\ve_0,\ldots,\ve_{N-1}$ in (\ref{eq:en}) as the gradients. Specifically, the $3N$-dimensional design-variable vector is defined by 
\begin{align*}
  \mathsf{x}:=\left((\vP_0)_x,(\vP_0)_y,(\vP_0)_z,\ldots,(\vP_{N-1})_x,(\vP_{N-1})_y,(\vP_{N-1})_z\right)^{\rm T}
\end{align*}
and the corresponding gradient vector is defined by
\begin{align*}
  \mathsf{g}:=\left((\ve_0)_x,(\ve_0)_y,(\ve_0)_z,\ldots,(\ve_{N-1})_x,(\ve_{N-1})_y,(\ve_{N-1})_z\right)^{\rm T},
\end{align*}
where $(\bm{f})_x$ stands for the $x$-component of a vector $\bm{f}$ and so on. Then, (\ref{eq:tilde J grad}) can be expressed as follows:
\begin{align*}
  K(\tilde{\mathsf{x}})=K(\mathsf{x})+\mathsf{g}(\mathsf{x})\cdot\delta\mathsf{x} + \Oee.
\end{align*}
The way to choose a vector $\delta\mathsf{x}$, that is, the update from a vector $\mathsf{x}$ at a certain step, depends on the choice of the solver for nonlinear programming. This study utilises the SLSQP (Sequential Least-Squares Quadratic Programming)~\cite{kraft1988software} available from the library NLopt~\cite{nlopt}. The user is required to provide a routine to evaluate $K$ and $\mathsf{g}$ for a given design vector $\mathsf{x}$.

\section{Results and discussion I: Verification}\label{s:sd_grad}

This section is dedicated to check the shape sensitivity $\ve_n$ derived in (\ref{eq:en}), which is the key of the proposed shape-optimisation method.

\subsection{Problem setting and the sensitivity}

Suppose a circular coil $C$ on the $xy$-plane. The coil $C$ has the radius of $a$ and the centre at O. Also, a steady current $I$ flows through $C$ in the counter-clockwise on the $xy$-plane. Then, let us compute the sensitivity of the objective function $K$ in (\ref{eq:J grad}) with respect to the radius $a$, i.e. $\frac{\diff K}{\diff a}$. Under this setting, when the target point $\vx$ is chosen as a point on the $z$-axis, one can represent the gradient of $\vB$ and thus $K$ in (\ref{eq:J grad single}) explicitly. Specifically, $\vB$ is exactly given for such a target point $\vx=(0,0,z)^{\rm T}$ as follows \cite{feynman1965flp}:
\begin{align*}
  \vB^{\rm exact}(z)=\frac{\mu I a^2}{2(z^2+a^2)^{3/2}}\vez,
\end{align*}
where $\vez:=(0,0,1)^{\rm T}$ denotes the orthonomal vector for the $z$-axis. Therefore, the exact expression of $K$ is calculated as follows:
\begin{align}
  K^{\rm exact}(z) = \frac{1}{2}\left(-\frac{3 \mu I a^2 z}{2(z^2+a^2)^{5/2}}-g\right)^2.
\end{align}
In this example, not the CPs but the radius $a$ of the circular coil $C$ is selected as the design variable. Then, the shape derivative of $K$ with respect to $a$ is calculated as follows:
\begin{align}
  \frac{\diff K^{\rm exact}(z)}{\diff a}
  = \frac{3 \mu I z a(2z^2-3a^2)}{2(z^2+a^2)^{7/2}} \left(\frac{3 \mu I a^2 z}{2(z^2+a^2)^{5/2}}+g\right).
  \label{eq:dKda_exact}
\end{align}

The corresponding numerical shape-sensitivity can be obtained by considering the variation of $a$. To this end, the perturbation of the $n$th CP $\vP_n$ needs to be related to the that of the radius $a$. Specifically, when $a$ changes to $\tilde{a}$, $\vP_n$ needs to change to 
\begin{align*}
  \tilde{\vP}_n = \frac{\tilde{a}}{a}\vP_n.
\end{align*}
Therefore, the perturbation of $\vP_n$ is obtained as
\begin{align*}
  \delta\vP_n := \tilde{\vP}_n - \vP_n = \frac{\tilde{a}-a}{a}\vP_n =\frac{\delta a}{a}\vP_n,
\end{align*}
where $\delta a:=\tilde{a}-a$ denotes the perturbation of $a$. Plugging this into (\ref{eq:tilde J grad}) yields 
\begin{align*}
  \tilde{K}=K+\left(\sum_{n=0}^{N-1}\frac{\ve_n\cdot\vP_n}{a}\right)\delta a + \Oee.
\end{align*}
The coefficient of $\delta a$ represents the sensitivity of $K$ with respect to $a$, i.e.
\begin{align}
  \frac{\diff K}{\diff a}
  =\sum_{n=0}^{N-1}\frac{\ve_n\cdot\vP_n}{a}.
  \label{eq:dKda}
\end{align}
The sensitivity in (\ref{eq:dKda}) will be compared with that in (\ref{eq:dKda_exact}).

\subsection{Results and discussion}

Suppose $\mu=1$ (which will be used in the next section), $I=1$, and the target point $\vx$ is placed at $(0,0,1)^{\rm T}$ on the $z$-axis. Then, the numerical sensitivity $\frac{\diff K}{\diff a}$ in (\ref{eq:dKda}) is compared with the exact one in (\ref{eq:dKda_exact}) for the radius $a$ from $0.10$ to $1.00$ every $0.01$.

The circular coil $C$ is generated with 32 CPs (regardless of the value of $a$) and the quadratic B-spline functions. Also, the number of the quadratic points per knot-interval is selected as 24.

\begin{figure}[hbt]
  \centering
  \begin{tabular}{cc}
    \includegraphics[width=.35\textwidth]{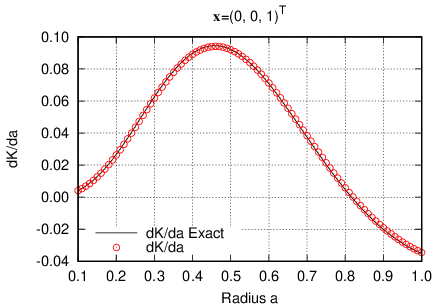}
    &\includegraphics[width=.35\textwidth]{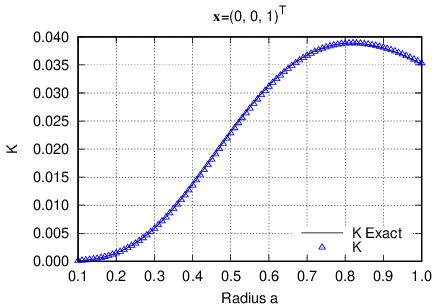}
  \end{tabular}
  \caption{Comparison of $\frac{\diff K}{\diff a}$ (left) and $K$ (right).}
  \label{fig:check_grad}
\end{figure}

Figure~\ref{fig:check_grad} compares the exact and numerical sensitivities in (\ref{eq:dKda_exact}) and (\ref{eq:dKda}) as well as those objective functions. Here, the target point $\vx$ was chosen as $(0,0,1)^{\rm T}$. Also, $g$ was selected as $0$ because the value of $g$ is unimportant for the present analysis. The figure shows a good agreement with regard to the sensitivity as well as the objective function. This result confirms the reliability of the proposed methodology based on the shape sensitivities.

\section{Results and discussion II: Shape optimisation}\label{s:gradcoil}

This section demonstrates some shape optimisations to obtain $z$-gradient coils.

\subsection{Problem setting}\label{s:gradcoil_problem}

\begin{figure}[hbt]
  \centering
  \includegraphics[width=.4\textwidth]{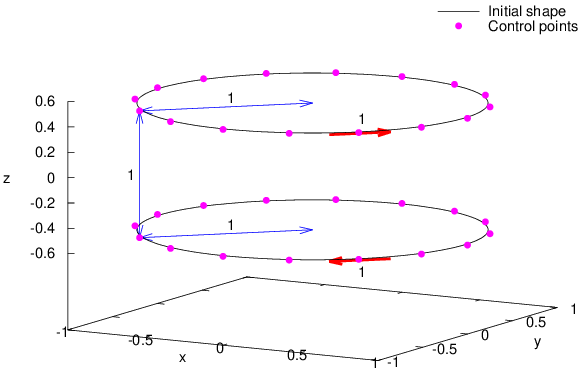}
  \caption{Problem setting of the shape optimisation. The red arrows show the direction of the steady currents of the magnitude of $1$.}
  \label{fig:gradcoil-problem-bird}
\end{figure}

\begin{figure}[hbt]
  \centering
  \begin{tabular}{cc}
    \multicolumn{2}{c}{\textbf{Case 1}: 11 target points}\\
    \includegraphics[width=.3\textwidth]{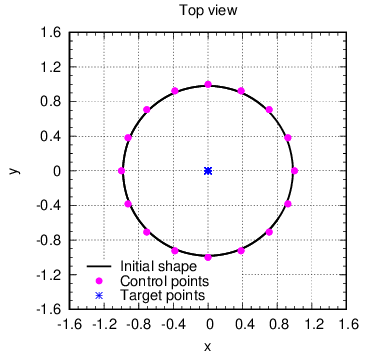}
    &\includegraphics[width=.3\textwidth]{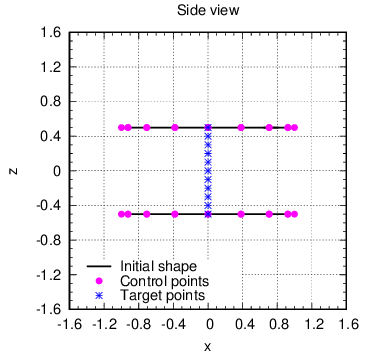}\\
    \multicolumn{2}{c}{\textbf{Case 2}: 55 target points}\\
    \includegraphics[width=.3\textwidth]{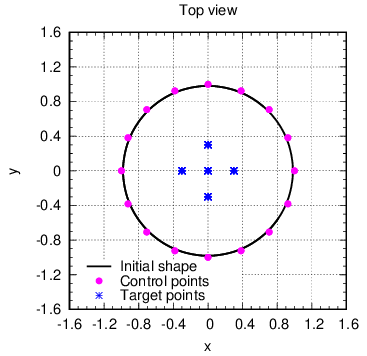}
    &\includegraphics[width=.3\textwidth]{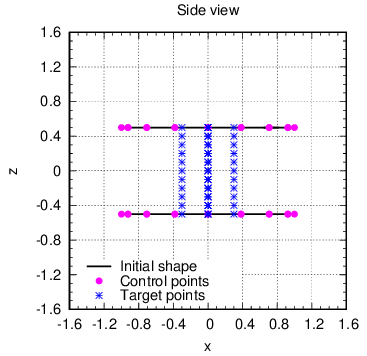}
  \end{tabular}
  \caption{Positions of target points.}
  \label{fig:gradcoil-cp-init}
\end{figure}

Let us consider a coil $C$ that consists of two closed curves. As shown in Figure~\ref{fig:gradcoil-problem-bird}, the curves are circles of radius $1$ and have their centres on the $z$-axis at the initial state. One circle is on the plane $z=-0.5$, while the other is on $z=0.5$. Steady state currents of the magnitude of $1$ are running in the opposite directions. Each coil is made of $16$ CPs and the quadratic B-spline functions. As the bounds of the CPs, every coordinate of each CP is allowed to change up to $\pm 0.3$ from the initial coordinate. This bound constraint is considered in order to prevent the coils from becoming highly complicated and thus to maintain the volume inside the coil to some extent. Then, it is considered to minimise the objective function $K$ in (\ref{eq:J grad general}). Regarding the number and position of the target points, the following two cases are considered as illustrated in Figure~\ref{fig:gradcoil-cp-init}: 
\begin{description}
\item[Case 1] 11 points on the $z$-axis: $(0.0,0.0,-0.5)$, $(0.0,0.0,-0.4)$, $\ldots$, $(0.0,0.0,0.5)$.
\item[Case 2] 55 points in total; every 11 points are on a vertical line passing through one of five points, i.e. $(-0.3,0.0)$, $(0.0,-0.3)$, $(0.0,0.0)$, $(0.0,0.3)$, and $(0.3,0.0)$.
\end{description}
In addition, all the target values $g_i$ are selected as $1$($=:g$) in each case, meaning that a uniform gradient of $B_z$ in the $z$-direction is designed.

As mentioned in Section~\ref{s:nlopt}, the SLSQP~\cite{kraft1988software} is employed as the gradient-based solver. For comparison, the COBYLA~\cite{powell2007}, which is a \textit{gradient-free method}, is also used. An optimisation is stopped when the relative change in $K$ to the previous step is $10^{-5}$ or less. For safety, the maximum number of optimisation steps is set to 1000.

\subsection{Results and discussion}\label{s:gradcoil result}

\nprounddigits{6}

\begin{figure}[hbt]
  \centering
  \begin{tabular}{cc}
    \textbf{Case 1} & \textbf{Case 2}\\
    \includegraphics[width=.35\textwidth]{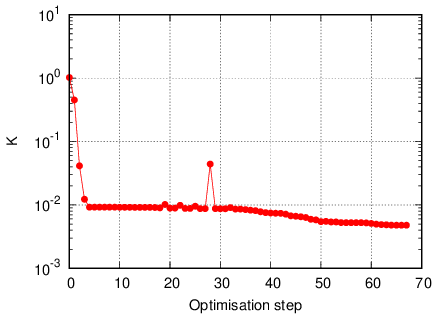}
    &\includegraphics[width=.35\textwidth]{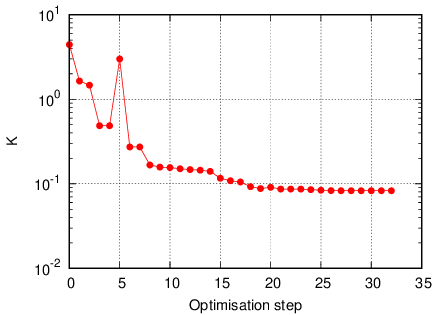}
  \end{tabular}
  \caption{History of the objective function $K$.}
  \label{fig:gradcoil-history}
\end{figure}

\begin{figure}[hbt]
  \centering
  \begin{tabular}{cc}
    \includegraphics[width=.35\textwidth]{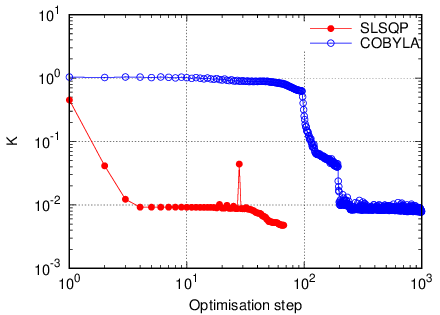}
    & \includegraphics[width=.3\textwidth]{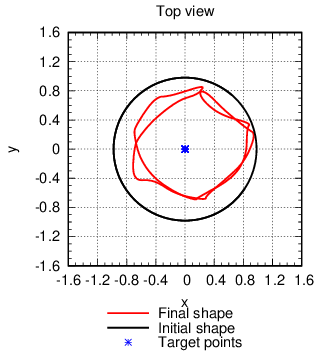}
  \end{tabular}
  \caption{(Left) Comparison of the gradient-based method (SLSQP) with the gradient-free method (COBYLA) in \textbf{Case 1}. It should be noted that the optimisation step is represented in the logarithmic scale. (Right) Final shape generated by the COBLYA.}
  \label{fig:gradcoil-comp}
\end{figure}

Figure~\ref{fig:gradcoil-history} shows that $K$ converged successfully in the both cases. For comparison, \textbf{Case 1} was performed with the COBYLA instead of the SLSQP. Then, as shown in Figure~\ref{fig:gradcoil-comp} (left), the optimisation did not stop before the predefined maximum steps, i.e. 1000. Further, the final shape in Figure~\ref{fig:gradcoil-comp} (right) was unreasonably asymmetric. This confirms that the shape derivative (sensitivities) is indispensable to solve a shape optimisation problem efficiently.

\begin{figure}[hbt]
  \centering
  \begin{tabular}{cc}
    \textbf{Case 1} & \textbf{Case 2}\\
    \includegraphics[width=.3\textwidth]{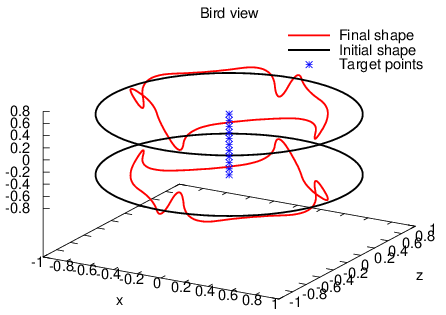}
    &\includegraphics[width=.3\textwidth]{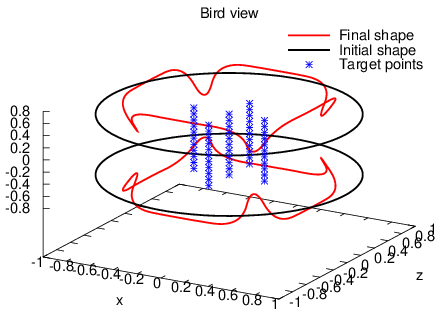}\\
    \includegraphics[width=.3\textwidth]{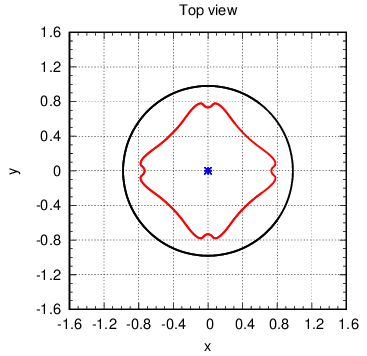}
    &\includegraphics[width=.3\textwidth]{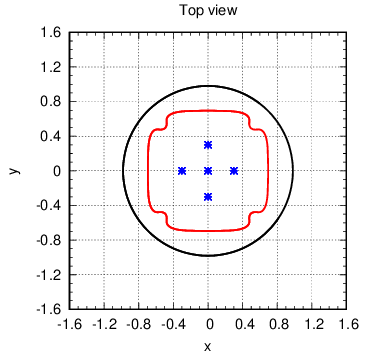}\\
    \includegraphics[width=.3\textwidth]{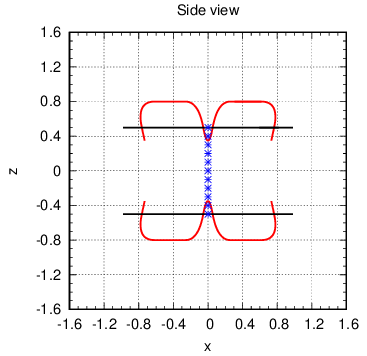}
    &\includegraphics[width=.3\textwidth]{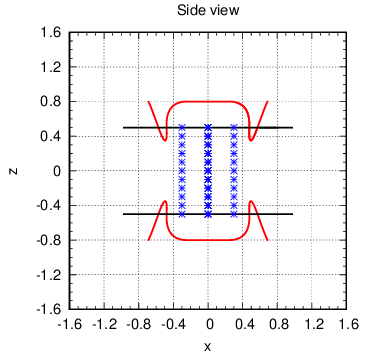}\\
  \end{tabular}
  \caption{Comparison of the optimised shape with the initial one.}
  \label{fig:gradcoil-shape}
\end{figure}

Figure~\ref{fig:gradcoil-shape} compares the optimised shape with the initial one for both cases. The two optimised shapes are similar to each other; when one is rotated by $45^\circ$ around the $z$-axis, the rotated shape resembles to the other. The optimised CPs are listed in Tables~\ref{tab:gradcoil-cp-final} and \ref{tab:gradcoil2-cp-final} of Section~\ref{s:cp}. From these tables, the $z$-coordinate of every CP is either the extrema $\pm 0.2$ or $\pm 0.8$, which means that all the CPs moved by the maximum amount, i.e. $0.3$, from the initial position in the $z$-direction.

\begin{figure}[hbt]
  \centering
  \begin{tabular}{cc}
    \textbf{Case 1} & \textbf{Case 2}\\
    \includegraphics[width=.35\textwidth]{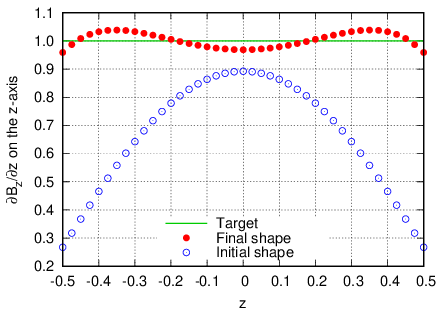}
    &\includegraphics[width=.35\textwidth]{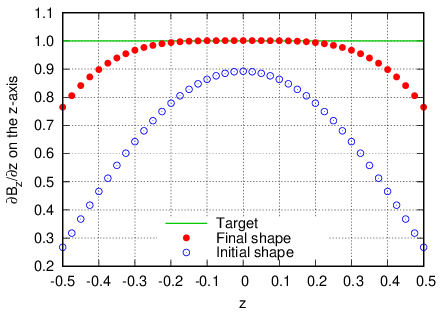}\\
    \includegraphics[width=.35\textwidth]{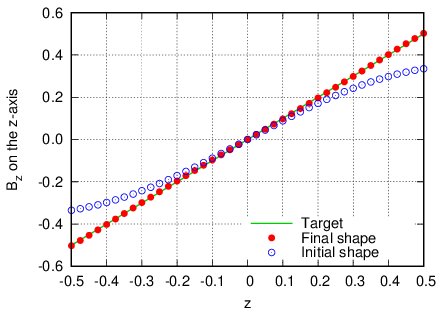}
    &\includegraphics[width=.35\textwidth]{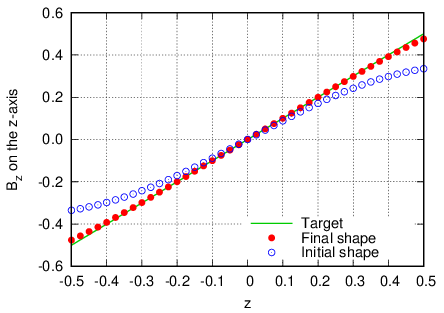}
  \end{tabular}
  \caption{Comparison of $\pp{B_z}{z}$ (top) and $B_z$ (bottom) on the $z$-axis.}
  \label{fig:gradcoil-dbzdz}
\end{figure}

\begin{figure}[hbt]
  \centering
  \begin{tabular}{ccc}
    Initial shape & \multicolumn{2}{c}{Final shape}\\
    & \textbf{Case 1} & \textbf{Case 2}\\
    \includegraphics[width=.3\textwidth]{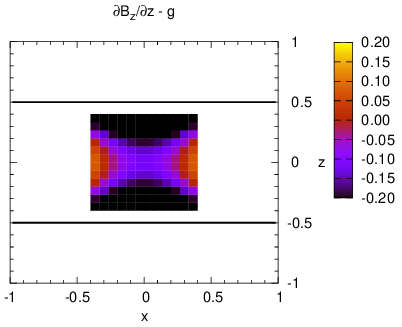}
    &\includegraphics[width=.3\textwidth]{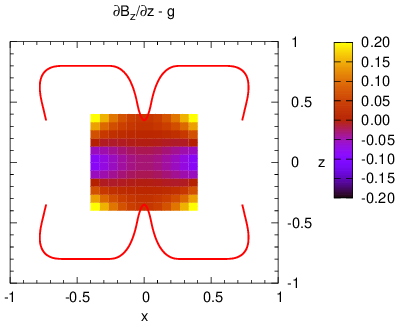}
    &\includegraphics[width=.3\textwidth]{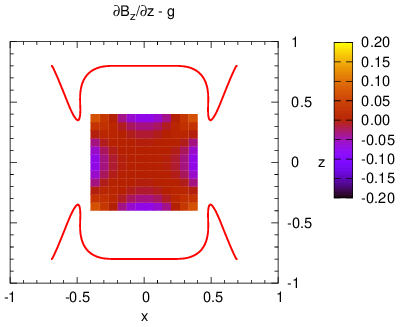}\\
    \includegraphics[width=.3\textwidth]{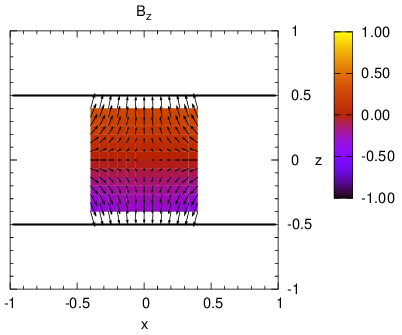}
    &\includegraphics[width=.3\textwidth]{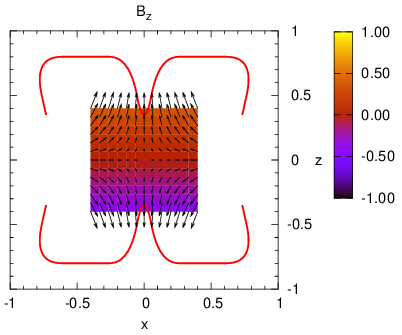}
    &\includegraphics[width=.3\textwidth]{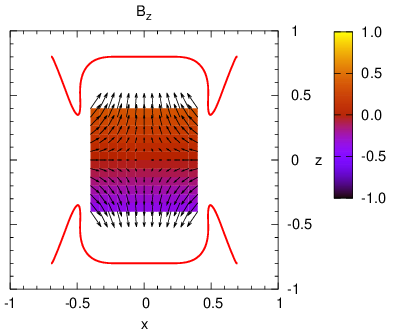}
  \end{tabular}
  \caption{(Top) Plot of $\pp{B_z}{z}-g$; the value is truncated to $\pm 0.2$ to clarify the difference from zero. (Bottom) Plot of $B_z$ together with $\bm{B}$ vectors  in the square region on the $xz$-plane, i.e. $\{(x,y,z)\ |\ -0.4\le x\le 0.4,\ y=0.0,\ -0.0\le z\le 0.4\}$. The displayed length of each vector is 15\% of the actual length for the sake of visibility.}
  \label{fig:gradcoil-pm3d}
\end{figure}

Figure~\ref{fig:gradcoil-dbzdz} shows $\pp{B_z}{z}$ and $B_z$ on the $z$-axis from $z=-0.5$ to $0.5$ every $0.02$. There is no significant difference in $B_z$ between the two cases. Regarding $\pp{B_z}{z}$, \textbf{Case 1} is close to the target value of $1$ in the region such as $|z|\le 0.5$, while \textbf{Case 2} is very close to $1$ in the relatively narrow region such as $|z|\le 0.2$ but highly different from the target value outside the region. In Figure~\ref{fig:gradcoil-pm3d}, the same features are observed in the square region on the $xz$-plane, i.e. $\{(x,y,z)\ |\ -0.4\le x\le 0.4,\ y=0.0,\ -0.0\le z\le 0.4\}$.

A more robust and longer linearly-changing $B_z$-field could be designed by, for example, increasing the number of CPs and/or coils, adjusting the maximum amount of translation of CPs, and arranging the target points. However, this study does not pursue it any more because the present results are satisfactory to manifest the validity, usability, and applicability of the proposed shape-optimisation method.

\subsection{Further discussions}\label{s:gradcoil further}

It would be valuable to discuss the initial conditions, i.e. the initial positions of CPs. They can affect the optimisation result. The principal reasons is that the employing nonlinear-programming solver SLSQP~\cite{kraft1988software} is to find not the global solution but the local solution. Another reason is to use the bound constraints, i.e. the allowable amount of moving of CPs from their initial positions.

\begin{figure}[hbt]
  \centering
  \includegraphics[width=.4\textwidth]{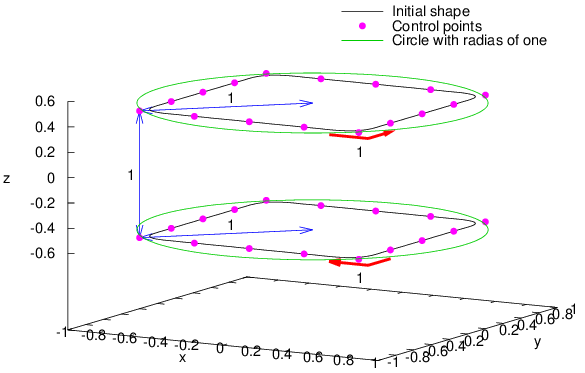}
  \caption{Problem setting of the shape optimisation for the rectangular coils at the initial step. The red arrows show the direction of the steady currents of the magnitude of $1$.}
  \label{fig:gradcoil5-problem-bird}
\end{figure}

To see the dependency of the result on the initial shape, square coils are considered instead of the circular ones at the initial shape (Figure~\ref{fig:gradcoil5-problem-bird}). The size of the squares is selected as $\sqrt{2}$ so that they are inscribed to the circles with radius of $1$. Each square coil is drawn with 16 CPs. Then, each square has rounded corners because any CPs are placed at the same position in this setting. In what follows, the 11 target points for \textbf{Case 1} is considered.

\begin{figure}[hbt]
  \centering
  \begin{tabular}{ccc}
    \includegraphics[width=.3\textwidth]{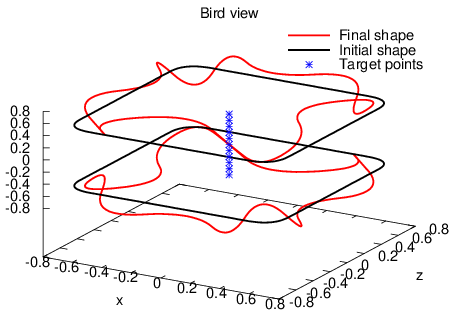}
    &\includegraphics[width=.3\textwidth]{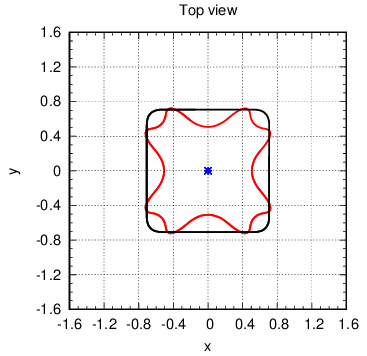}    
    &\includegraphics[width=.3\textwidth]{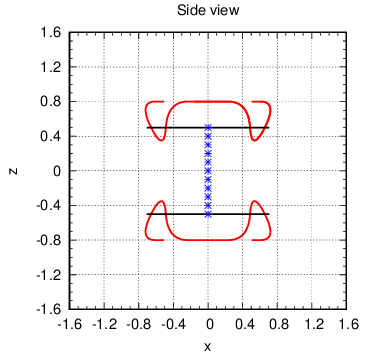}\\
    \includegraphics[width=.3\textwidth]{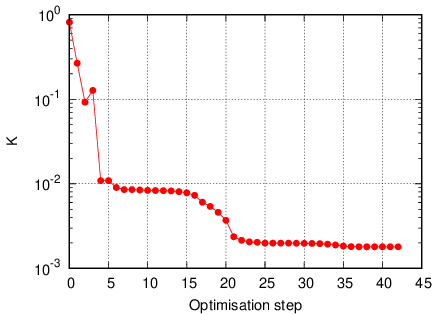}
    &\includegraphics[width=.3\textwidth]{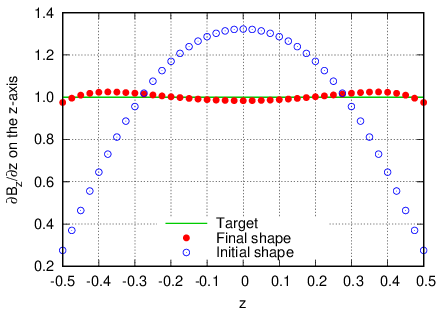}
    &\includegraphics[width=.3\textwidth]{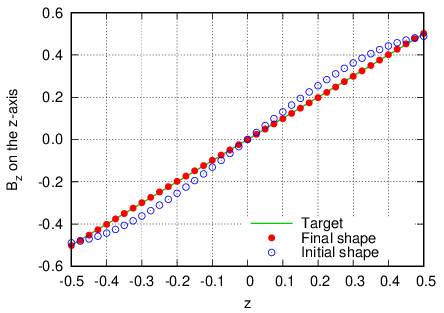}
  \end{tabular}
  \caption{Results of the shape optimisation with using the rectangular coils at the initial step.}
  \label{fig:gradcoil5-result}
\end{figure}

Figure~\ref{fig:gradcoil5-result} shows the results. The optimisation ended at the 42nd step. The present square case achieved smaller value of $K$ than the circular case shown in Figure~\ref{fig:gradcoil-history}. This is actually observed in the plot of $\pp{B_z}{z}$ when it is compared with that of Figure~\ref{fig:gradcoil-dbzdz}. However, it is difficult to say that the square case is definitely better than the circular case because the initial distribution of $B_z$ on the $z$-axis of the square case is more close to the target distribution, which is $B_z(\vx)=z$ in the present case, than that of the circular case, which is plotted in Figure~\ref{fig:gradcoil-dbzdz}. So, if the initial radius was given as a smaller value than one (without changing the bound constraints of CPs), the circular case could outperform the square case.

Another interesting aspect of the proposed optimisation method is the applicability to multiple-turn coils, which are usually used for MRI. Handling such coils is principally possible, but many turns can consume a lot of CPs and thus the resulting nonlinear programming problem can be a large-scale problem that is computationally expensive. In addition, the constraints of the positions of CPs can become more important because the many and/or longer coils can be intersected one another. Such intersections are mathematically no problem in the present framework as long as any target point, where the magnetic field is evaluated, is not on any coil in the optimisation process. However, fabricating intersected coils is physically impossible. It is thus necessary to slightly modify the trajectory of the optimised coils, but such a modification could degrade the original performance. Hence, handling multiple-turn coils is challenging, but the present study can be the starting point.

\begin{figure}[hbt]
  \centering
  \includegraphics[width=.6\textwidth]{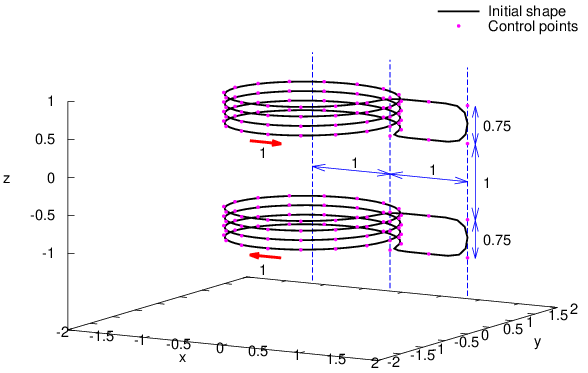}
  \caption{Problem setting of the shape optimisation for two 4-turn coils. The red arrows show the direction of the steady currents of the magnitude of 1.}
  \label{fig:gradcoil6-problem-bird}
\end{figure}

\begin{figure}[hbt]
  \centering
  \begin{tabular}{ccc}
    \includegraphics[width=.35\textwidth]{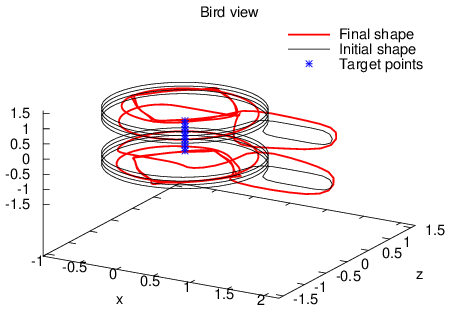}
    &\includegraphics[width=.275\textwidth]{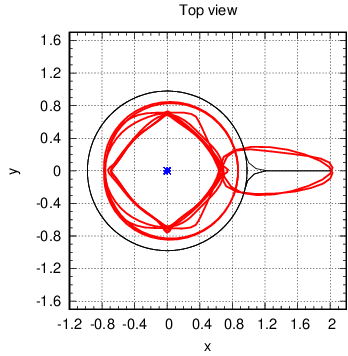}    
    &\includegraphics[width=.275\textwidth]{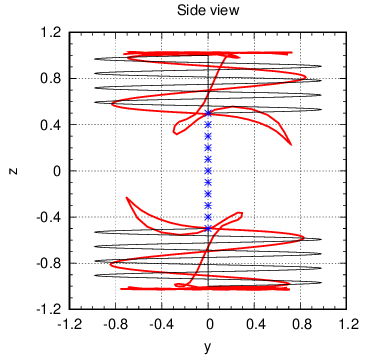}\\
    \includegraphics[width=.3\textwidth]{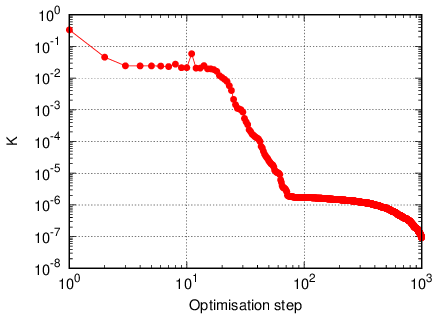}
    &\includegraphics[width=.3\textwidth]{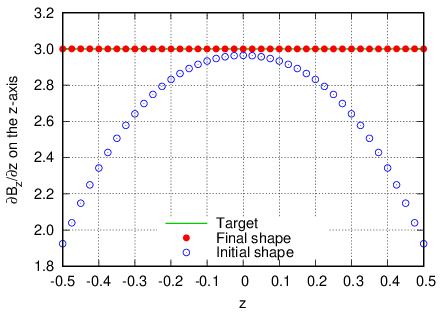}
    &\includegraphics[width=.3\textwidth]{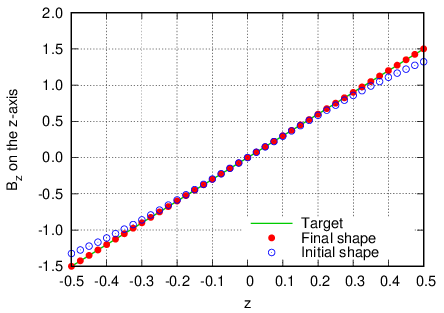}
    \end{tabular}
  \caption{Results of the shape optimisation with using the 4-turn coils at the initial step.}
  \label{fig:gradcoil6-result}
\end{figure}

To see the capability of the present coil-optimisation method, 4-turn coils are considered in a similar setting of the present optimisation problem. Basically, the circular coils (recall Figure~\ref{fig:gradcoil-problem-bird}) are replaced with 4-turn coils, which are illustrated in Figure~\ref{fig:gradcoil6-problem-bird}. Each coil consists of a helical part (radius of $1$; height of $0.75$) and a connecting line, which is about $2.75$ length and joins the ends of the helix. Each coil is constructed with $69$ CPs and thus the total number of design variables is $414$. Again, each CP can move up to $\pm 0.3$ in every direction. This could induce intersections of coils, but no workaround is considered in this optimisation. Regarding all the target values of the $z$-gradient (i.e. $g_i$ in $K$) are uniformly set to $3$ (instead of $1$) because the initial gradient along the $z$-axis is close to $3$ (rather than $1$), which will be seen in Figure~\ref{fig:gradcoil6-result}.

Figure~\ref{fig:gradcoil6-result} shows the result of the 4-turn coils. As shown in the history of $K$, the optimisation using the SLSQP~\cite{kraft1988software} stopped due to the limited number of optimisation steps, i.e. 1000. However, the present behaviour of the history is different from that of the COBYLA~\cite{powell2007} in the case of the circular coils (recall Figure~\ref{fig:gradcoil-comp}). Actually, the former is still decreasing at the final step, while the latter saturates after about the 250th steps. In addition, the final shape of the present optimisation in Figure~\ref{fig:gradcoil6-result} reveals a symmetry with respect to the $x$-axis. It is difficult to explain why the final shape takes such a form, but the optimised gradient significantly agrees with the target one, which is observed in the sub-figure for $\pp{B_z}{z}$. In summary, although the stopping criterion needs to be modified so that very small value of $K$ can be discarded, it was demonstrated that the present optimisation for not many-turn but non-single-turn coils was successfully performed with the proposed method.

\section{Conclusion}\label{s:conclusion}

The present study established a new method to directly optimise the shape of gradient coils so that $\pp{B_z}{z}$ (i.e. the $z$-gradient of the $z$-component of the magnetic field $\bm{B}$) can be agreed with given target values. The core of the proposed method is the derivation of the shape derivative $E$ in (\ref{eq:J grad sd}) for the objective function $K$ in (\ref{eq:J grad}). The shape derivative $E$ is discretised with the closed B-spline curves to obtain the shape sensitivities of $K$ with respect to the control points (CPs) of the curves. These sensitivities can be regarded as the first order gradient in terms of nonlinear programming. Therefore, the shape optimisation problems of interest can be solved by means of a general-purpose solver for nonlinear programming. After a verification of $E$ in Section~\ref{s:sd_grad}, the shape optimisations for some $z$-gradient coils were demonstrated and discussed in Section~\ref{s:gradcoil}. The results show that the proposed shape-optimisation method is promising for designing gradient coils regarding MRI.

There are some future tasks in addition to developing an efficient way to handle multiple-turn coils, which was discussed in Section~\ref{s:gradcoil further}. First, in order to realise a rapid gradient switching \cite{hidalgo2010}, the inductance needs to be considered as an objective function. The corresponding shape derivative can be derived in the same way as the present paper, but the details will be reported in the subsequent paper. Second, when concerning about the heat of coils, it is desirable to restrict the length of coils. The length can be treated as a constraint of shape optimisation. In fact, the length of a coil can be described with a function of design variables, that is, CPs. Moreover, reducing acoustic noise in MR systems (e.g. functional MRI~\cite{palmer2006new_fMRI}) is crucial for the patients and the environments \cite{motovilova2022}. For this regard, the Lorentz force acting on a gradient coil should be taken into account as a part of objective function \cite{mansfield1995}. With all these future developments, the proposed design method would become more versatile and valuable for practical designs of gradient coils.

\appendix

\section{Derivation of (\ref{eq:J grad update})}\label{s:derive_sd_grad}

Correspondingly to $\tilde{\vs}$, the perturbed magnetic field $\tilde{\vB}=\tilde{\vB}(\vx)$ is represented as
\begin{align}
  \tilde{\vB}=\frac{\mu I}{4\pi}\int_{\tilde{C}}\frac{\diff\tilde{\vs}\times(\vx-\tilde{\vs})}{|\vx-\tilde{\vs}|^3}.
  \label{eq:tilde B tmp}
\end{align}
From (\ref{eq:tilde s}) and its direct consequence
\begin{align}
  \diff\tilde{\vs}:=\diff\vs+\epsilon\diff\vV(\vs),
  \label{eq:diff tilde s}
\end{align}
(\ref{eq:tilde B tmp}) can be written as
\begin{align*}
  \tilde{\vB}=\frac{\mu I}{4\pi}\int_C\frac{\left(\diff\vs+\epsilon\diff\vV\right)\times\left(\vx-\vs-\epsilon\vV\right)}{|\vx-\vs-\epsilon\vV|^3}.
\end{align*}
Here, the numerator in the RHS is calculated as
\begin{align*}
  \left(\diff\vs+\epsilon\diff\vV\right)\times\left(\vx-\vs-\epsilon\vV\right)
  =\diff\vs\times(\vx-\vs)+\epsilon\{-\diff\vs\times\vV+\diff\vV\times(\vx-\vs)\}+\Oee.
\end{align*}
On the other hand, the denominator is calculated as 
\begin{align*}
  \frac{1}{\left[(\vx-\vs-\epsilon\vV)\cdot(\vx-\vs-\epsilon\vV)\right]^{\frac{3}{2}}}
  &=\frac{1}{\left[(\vx-\vs)^2-2\vV\cdot(\vx-\vs)\epsilon+\epsilon^2\vV\cdot\vV\right]^{\frac{3}{2}}}\nonumber\\
  &=\frac{1}{|\vx-\vs|^3\left[1-\frac{2\vV\cdot(\vx-\vs)}{|\vx-\vs|^2}\epsilon+\Oee\right]^{\frac{3}{2}}}\nonumber\\
  &=\frac{1}{|\vx-\vs|^3}\left[1+3\frac{\vV\cdot(\vx-\vs)}{|\vx-\vs|^2}\epsilon\right]+\Oee.
\end{align*}
Therefore, $\tilde{\vB}$ can be expressed as
\begin{align}
  \tilde{\vB}
  &=\frac{\mu I}{4\pi}\int_C
  \left[\diff\vs\times(\vx-\vs)+\epsilon\{-\diff\vs\times\vV+\diff\vV\times(\vx-\vs)\}\right]\times\left(\frac{1}{|\vx-\vs|^3}+\epsilon\frac{3\vV\cdot(\vx-\vs)}{|\vx-\vs|^5}\right) + \Oee \nonumber\\
  &=\frac{\mu I}{4\pi}\int_C\frac{\diff\vs\times(\vx-\vs)}{|\vx-\vs|^3}+\epsilon\frac{\mu I}{4\pi}\int_C\left[\frac{\diff\vs\times(\vx-\vs)}{|\vx-\vs|^5}3\vV\cdot(\vx-\vs)+\frac{-\diff\vs\times\vV+\diff\vV\times(\vx-\vs)}{|\vx-\vs|^3}\right]+\Oee \nonumber\\
  &=\vB+\epsilon\frac{\mu I}{4\pi}\int_C\left[\frac{3\diff\vs\times\vr}{r^5}\vV\cdot\vr+\frac{\diff\vV\times\vr-\diff\vs\times\vV}{r^3}\right]+\Oee,
  \label{eq:tilde B}
\end{align}
where $\vr:=\vx-\vs$ and $r:=|\vr|$ are defined. Hence, the $z$-component of $\tilde{\vB}$ in (\ref{eq:tilde B}) is described as follows:
\begin{align*}
  \tilde{B}_z = B_z+\epsilon\frac{\mu I}{4\pi}\int_C\left[\frac{3(\diff\vs\times\vr)_z}{r^5}\vV\cdot\vr+\frac{(\diff\vV\times\vr-\diff\vs\times\vV)_z}{r^3}\right].
\end{align*}
Here, it should be recalled that $\vr:=\vx-\vs=(r_x,r_y,r_z)^{\rm T}=(x-s_x,y-s_y,z-s_z)^{\rm T}$. Hence, $(\diff\vs\times\vr)_z=\diff s_x (y-s_y) - \diff s_y (x-s_x)$ is independent of $z$. This is true also for $(\diff\vV\times\vr-\diff\vs\times\vV)_z$. Therefore, the differentiation of $\tilde{B}_z$ with respect to $z$ can be written as follows:
\begin{align*}
  \pp{\tilde{B}_z}{z}
  =& \pp{B_z}{z}+\epsilon\frac{\mu I}{4\pi}\int_C\left[3(\diff\vs\times\vr)_z\vV\cdot\frac{\partial}{\partial z}\left(\frac{\vr}{r^5}\right)+(\diff\vV\times\vr-\diff\vs\times\vV)_z\frac{\partial}{\partial z}\left(\frac{1}{r^3}\right)\right] + \Oee.
\end{align*}
Since, letting $\vez$ be the unit vector $(0,0,1)^{\rm T}$, 
\begin{align*}
  \frac{\partial}{\partial z}\left(\frac{\vr}{r^5}\right)
  &=\frac{\pp{\vr}{z}r^5-\vr\pp{r^5}{z}}{r^{10}}=\frac{\vez}{r^5}-\frac{5r_z\vr}{r^7}\\
  \intertext{and}
  \frac{\partial}{\partial z}\left(\frac{1}{r^3}\right)
  &=-3r^{-4}\pp{r}{z}=-\frac{3r_z}{r^5}
\end{align*}
hold, one gets
\begin{align*}
  \pp{\tilde{B}_z}{z}
  = \pp{B_z}{z}+\epsilon\frac{\mu I}{4\pi}\int_C\left[3(\diff\vs\times\vr)_z\left(\frac{V_z}{r^5}-\frac{5r_z}{r^7}\vV\cdot\vr\right)-\frac{3r_z(\diff\vV\times\vr-\diff\vs\times\vV)_z}{r^5}\right] + \Oee.
\end{align*}
Subtracting $g$ from the both sides and then squaring the resulting expression yield
\begin{align*}
  \left(\pp{\tilde{B}_z}{z} - g \right)^2
  = \left(\pp{B_z}{z} - g \right)^2
  +2\left(\pp{B_z}{z}-g\right)\epsilon\frac{\mu I}{4\pi}\int_C\left[3(\diff\vs\times\vr)_z\left(\frac{V_z}{r^5}-\frac{5r_z}{r^7}\vV\cdot\vr\right)-\frac{3r_z(\diff\vV\times\vr-\diff\vs\times\vV)_z}{r^5}\right] + \Oee.
\end{align*}
This is exactly the desired equation in (\ref{eq:J grad update}) because of the definition of $K$ in (\ref{eq:J grad single}).

\section{Construction of closed B-spline curves}\label{s:bspline}

This study constructs a closed B-spline curve by wrapping knots~\cite{hoschek1993}. To do so, a set of knots or a \textit{knot vector} is introduced as
\begin{align*}
  T:=\{t_0,t_1,\ldots,t_N,t_{N+1},\ldots,t_{N+p}\},
\end{align*}
where the inequalities
\begin{align}
  0=t_0 < t_1 < \cdots <t_N = 1 \label{eq:nonover}
\end{align}
are assumed for the first $N+1$ knots and the remaining $p$ knots, where $p$ stands for the degree of the B-spline functions mentioned below, are defined by
\begin{align}
  t_i := t_N+t_{i-N}-t_0\quad\text{for $i=N+1,\ldots,N+p$} \label{eq:wrap}.
\end{align}
Here, (\ref{eq:nonover}) means that knots can be non-uniform but non-overlapped and (\ref{eq:wrap}) is said that the last $p$ knots are wrapped to the first ones. 

The usual B-spline functions, defined with $T$ and denoted by $N_0^p$, $\ldots$, $N_{N-1}^p$, do not satisfy the partition of unity (POU) on the domain $[t_0,t_N]$, i.e. $\sum_{n=0}^{N-1}N_j^p(t)\ne 1$, for any $t\in[t_0,t_N]$. Thus, they cannot be used to construct a closed curve. To satisfy it, for any B-spline function, the part corresponding to $t>t_N$ may be shifted horizontally by $-t_N$. In other words, the following \textit {periodic B-spline function} $R_j^p$ may be used instead of $N_j^p$ itself:
\begin{align*}
  R_j^p(t) := N_j^p(t) + N_j^p(t+t_N)\quad\text{for $j=0,\ldots,N-1$}.
\end{align*}
For example, $R_j^2$ is expressed as follows:
\begin{align*}
  R_j^2(t)=&\left\{
  \begin{array}{ll}
    0 & t\le t_{j}\\
    \frac{(t-t_j)^2}{(t_{j+2}-t_j)(t_{j+1}-t_j)} & t\le t_{j+1}\\
    \frac{(t-t_j)(t_{j+2} - t)}{(t_{j+2} - t_j)(t_{j+2} - t_{j+1})}+\frac{(t-t_{j+1})(t_{j+3} - t)}{(t_{j+3} - t_{j+1}) (t_{j+2} - t_{j+1})} & t\le t_{j+2}\\
    \frac{(t_{j+3} - t)^2}{(t_{j+3} - t_{j+1}) (t_{j+3} - t_{j+2})} & t\le t_{j+3}\\
    0 & \text{elsewhere}
  \end{array}\right. && \text{for $0\le j\le N-3$},\\
  R_{N-2}^2(t)=&\left\{
  \begin{array}{ll}
    \frac{(t_{1} - t)^2}{(t_{1} - (t_{N-1} - t_{N}))  (t_{1} - t_{0})} & t \le t_{1}\\
    0 & t \le t_{N-2}\\
    \frac{(t - t_{j})^2}{(t_{j+2} - t_{j})  (t_{j+1} - t_{j})} & t \le t_{N-1}\\
    \frac{(t-t_{j})(t_{j+2}) - t)}{(t_{j+2} - t_{j})  (t_{j+2} - t_{j+1})}+\frac{(t-t_{j+1})(t_{j+3}-t)}{(t_{j+3} - t_{j+1})  (t_{j+2} - t_{j+1})} & t \le t_{N}
  \end{array}
  \right.,\\
  R_{N-1}^2(t)=&\left\{
  \begin{array}{ll}
    \frac{t-(t_{N-1}-t_{N}))(t_{1} - t)}{(t_{1} - (t_{N-1}-t_{N}))  (t_{1} - t_{0})}+\frac{(t-t_{0})(t_{2}-t)}{(t_{2} - t_{0})  (t_{1} - t_{0})} & t \le t_{1}\\
    \frac{(t_{2} - t)^2}{(t_{2} - t_{0})  (t_{2} - t_{1})} & t \le t_{2}\\
    0 & t \le t_{N-1}\\
    \frac{(t - t_{N-1})^2}{(t_{n+1} - t_{N-1})  (t_{N} - t_{N-1})} & t \le t_{N}
  \end{array}\right..
\end{align*}

With the periodic B-spline functions $R_n^p$, (\ref{eq:s with R}) gives a closed B-spline curve defined on the domain $[t_0,t_N](\equiv[0,1])$.

\begin{figure}[hbt]
  \centering
  \includegraphics[width=.7\textwidth]{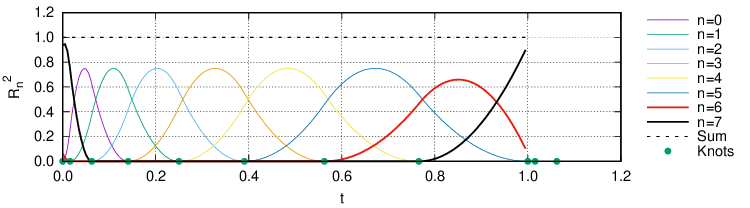}
  \caption{Non-uniform periodic B-spline functions $R_0^2$, $\ldots$, $R_7^2$ of $p=2$ and $N=8$. It is observed that $R_6^2$ and $R_7^2$ are split to two parts. The graph with 'Sum' denotes the summation $\sum_{n=0}^{N-1}R_j^2(t)$ to check the partition of unity.}
  \label{fig:bspline}
\end{figure}

Figure~\ref{fig:bspline} illustrates an example of non-uniform periodic B-spline functions when using $N=8$, $p=2$, and $t_i:=(i/N)^2$ for $i=0,\ldots,N$. It can be seen that $R_6^2$ and $R_7^2$ are partially shifted by $-1(\equiv -t_N)$. In addition, the graph labelled with 'Sum' in the figure denotes the summation $\sum_{n=0}^{N-1}R_j^2(t)$. It is observed that the set $\{R_0^2,\ldots,R_{N-1}^2\}$ satisfies the POU.

\section{Integration over $C$}\label{s:integral}

The contour integral over $C$ is first split to $N$ intervals, which were introduced in Section~\ref{s:closed B-spline}, and then evaluated for every interval by using $G$-points Gauss--Legendre quadrature method \cite{abramowitz1972} as follows:
\begin{align*}
  \int_C f(t) \diff t
  &= \sum_{k=0}^{N-1} \int_{I_k} f(t) \diff t 
  &&\reason{$C\equiv I_0\cup\cdots\cup I_{N-1}$}\nonumber\\
  &= \sum_{k=0}^{N-1} \int_{-1}^1 f\left(\frac{1-\xi}{2}t_k+\frac{1+\xi}{2}t_{k+1}\right) \frac{t_{k+1}-t_k}{2}\diff\xi 
  &&\reason{variable transformation from $t$ to $\xi$} \nonumber\\
  &\approx \sum_{k=0}^{N-1} \sum_{g=1}^{G}f\left(\frac{1-\xi_g}{2}t_k+\frac{1+\xi_g}{2}t_{k+1}\right) \frac{t_{k+1}-t_k}{2}w_g,
\end{align*}
where $\xi_g\in[-1,1]$ and $w_g$ denote the $g$th quadrature point and weight, respectively.

\begin{remark}
  If a coil $C$ consist of multiple closed curves, say $C_1$, $\ldots$, $C_M$, one may split the contour integral over $C$ as those of closed curves, i.e. $\int_C = \sum_{m=1}^M \int_{C_i}$. Here, each closed curve has its own curve parameter, knots, and the (periodic) B-spline functions.
\end{remark}

\section{Derivation of (\ref{eq:tilde J grad})}\label{s:derive_sensitivity_grad}

The substitution of $\epsilon\vV$ in (\ref{eq:eV}) into $K(\tilde{C})$ in (\ref{eq:J grad update}) yields
\begin{align*}
  K(\tilde{C})
  =K(C)+\left(\pp{B_z}{z}-g\right)\frac{\mu I}{4\pi}\int_C\Biggl[&
    3(\diff\vs\times\vr)_z\left(\frac{1}{r^5}\sum_n R_n(\delta\vP_n)_z-\frac{5r_z}{r^7}\sum_nR_n\delta\vP_n\cdot\vr\right)\nonumber\\
    &-\frac{3r_z}{r^5}\left(\sum_n\diff R_n\delta\vP_n\times\vr-\diff\vs\times\sum_nR_n\delta\vP_n\right)_z\Biggr] + \Oee \nonumber\\
  =K(C)+\sum_n\left(\pp{B_z}{z}-g\right)\frac{\mu I}{4\pi}\int_C\Biggl[&
    \underbrace{\frac{3(\dot{\vs}\times\vr)_z R_n(\delta\vP_n)_z}{r^5}}_{=:\alpha}-\underbrace{\frac{15(\dot{\vs}\times\vr)_z r_z R_n\delta\vP_n\cdot\vr}{r^7}}_{=:\beta}\nonumber\\
    &-\underbrace{\frac{3r_z\left(\dot{R}_n\delta\vP_n\times\vr-\dot{\vs}\times R_n\delta\vP_n\right)_z}{r^5}}_{=:\gamma}\Biggr] \diff t+ \Oee.
\end{align*}
Here, the terms $\alpha$, $\beta$, and $\gamma$ can be calculated as
\begin{align*}
  \alpha&=\frac{3}{r^5}\begin{pmatrix}0 \\ 0 \\ (\dot{\vs}\times\vr)_zR_n\end{pmatrix}\cdot\delta\vP_n, \nonumber\\
    \beta&=\frac{15(\dot{\vs}\times\vr)_z r_z R_n}{r^7}\vr\cdot\delta\vP_n, \nonumber\\
    \gamma&=\frac{3r_z}{r^5}\left\{
    \dot{R}_n\left((\delta\vP_n)_x r_y - (\delta\vP_n)_y r_x\right)
    - R_n\left(\dot{s}_x(\delta\vP_n)_y - \dot{s}_y(\delta\vP_n)_x\right)
    \right\}
    =\frac{3r_z}{r^5}
    \begin{pmatrix}
      \dot{R}_n r_y + R_n \dot{s}_y \\
      - \dot{R}_n r_x - R_n \dot{s}_x\\
      0
    \end{pmatrix}
    \cdot\delta\vP_n.
\end{align*}
Therefore, one can obtain
\begin{align*}
  K(\tilde{C})
  = K(C) +\sum_n\left(\pp{B_z(\vx)}{z}-g\right)\frac{\mu I}{4\pi}\int_C
  \left[\alpha-\beta-\gamma\right]\diff t + \Oee.
\end{align*}
The calculation of $\alpha-\beta-\gamma$ in the RHS derives the desired expression of $\tilde{K}=K(\tilde{C})$ in (\ref{eq:tilde J grad}) together with the shape sensitivity $\ve_n$ in (\ref{eq:en}).

\section{Tables of the optimised control points in Section~\ref{s:gradcoil}}\label{s:cp}

Tables~\ref{tab:gradcoil-cp-final} and \ref{tab:gradcoil2-cp-final} show the optimised control points computed in Section~\ref{s:gradcoil}. As pointed out in the section, the $z$-coordinate of every CP is either $\pm 0.2$ or $\pm 0.8$ in the both cases.

\begin{table}[hbt]
  \centering
  \caption{List of the 32 control points for the optimised shape in \textbf{Case 1}.}
  \label{tab:gradcoil-cp-final}
  \begin{tabular}{cc}
    Lower coil & Upper coil\\
    \begin{tabular}{ccc}
      \hline
      $x$ & $y$ & $z$\\ 
      \hline
      7.000000e-01 & -7.020514e-14 & -2.000000e-01\\
      8.282418e-01 & -1.085038e-01 & -8.000000e-01\\
      4.071068e-01 & -4.071068e-01 & -8.000000e-01\\
      1.085038e-01 & -8.282418e-01 & -8.000000e-01\\
      -5.082835e-14 & -7.000000e-01 & -2.000000e-01\\
      -1.085038e-01 & -8.282418e-01 & -8.000000e-01\\
      -4.071068e-01 & -4.071068e-01 & -8.000000e-01\\
      -8.282418e-01 & -1.085038e-01 & -8.000000e-01\\
      -7.000000e-01 & -4.536269e-14 & -2.000000e-01\\
      -8.282418e-01 & 1.085038e-01 & -8.000000e-01\\
      -4.071068e-01 & 4.071068e-01 & -8.000000e-01\\
      -1.085038e-01 & 8.282418e-01 & -8.000000e-01\\
      2.897734e-15 & 7.000000e-01 & -2.000000e-01\\
      1.085038e-01 & 8.282418e-01 & -8.000000e-01\\
      4.071068e-01 & 4.071068e-01 & -8.000000e-01\\
      8.282418e-01 & 1.085038e-01 & -8.000000e-01\\
      \hline
    \end{tabular}
    &
    \begin{tabular}{ccc}
      \hline
      $x$ & $y$ & $z$\\ 
      \hline
      7.000000e-01 & 1.142005e-13 & 2.000000e-01\\
      8.282418e-01 & 1.085038e-01 & 8.000000e-01\\
      4.071068e-01 & 4.071068e-01 & 8.000000e-01\\
      1.085038e-01 & 8.282418e-01 & 8.000000e-01\\
      -8.278121e-14 & 7.000000e-01 & 2.000000e-01\\
      -1.085038e-01 & 8.282418e-01 & 8.000000e-01\\
      -4.071068e-01 & 4.071068e-01 & 8.000000e-01\\
      -8.282418e-01 & 1.085038e-01 & 8.000000e-01\\
      -7.000000e-01 & 3.369748e-14 & 2.000000e-01\\
      -8.282418e-01 & -1.085038e-01 & 8.000000e-01\\
      -4.071068e-01 & -4.071068e-01 & 8.000000e-01\\
      -1.085038e-01 & -8.282418e-01 & 8.000000e-01\\
      -4.242781e-15 & -7.000000e-01 & 2.000000e-01\\
      1.085038e-01 & -8.282418e-01 & 8.000000e-01\\
      4.071068e-01 & -4.071068e-01 & 8.000000e-01\\
      8.282418e-01 & -1.085038e-01 & 8.000000e-01\\
      \hline
    \end{tabular}
  \end{tabular}
\end{table}

\begin{table}[hbt]
  \centering
  \caption{List of the 32 control points for the optimised shape in \textbf{Case 2}.}
  \label{tab:gradcoil2-cp-final}
  \begin{tabular}{cc}
    Lower coil & Upper coil\\
    \begin{tabular}{ccc}
      \hline
      $x$ & $y$ & $z$\\ 
      \hline
      7.000000e-01 & -4.615890e-14 & -8.000000e-01\\
      6.796808e-01 & -4.914970e-01 & -8.000000e-01\\
      4.645086e-01 & -4.645086e-01 & -2.000000e-01\\
      4.914970e-01 & -6.796808e-01 & -8.000000e-01\\
      -1.596331e-13 & -7.000000e-01 & -8.000000e-01\\
      -4.914970e-01 & -6.796808e-01 & -8.000000e-01\\
      -4.645086e-01 & -4.645086e-01 & -2.000000e-01\\
      -6.796808e-01 & -4.914970e-01 & -8.000000e-01\\
      -7.000000e-01 & -6.607883e-14 & -8.000000e-01\\
      -6.796808e-01 & 4.914970e-01 & -8.000000e-01\\
      -4.645086e-01 & 4.645086e-01 & -2.000000e-01\\
      -4.914970e-01 & 6.796808e-01 & -8.000000e-01\\
      -1.664327e-13 & 7.000000e-01 & -8.000000e-01\\
      4.914970e-01 & 6.796808e-01 & -8.000000e-01\\
      4.645086e-01 & 4.645086e-01 & -2.000000e-01\\
      6.796808e-01 & 4.914970e-01 & -8.000000e-01\\
      \hline
    \end{tabular}
    &
    \begin{tabular}{ccc}
      \hline
      $x$ & $y$ & $z$\\ 
      \hline
      7.000000e-01 & 1.136383e-13 & 8.000000e-01\\
      6.796308e-01 & 4.913340e-01 & 8.000000e-01\\
      4.646377e-01 & 4.646377e-01 & 2.000000e-01\\
      4.913340e-01 & 6.796308e-01 & 8.000000e-01\\
      2.369916e-13 & 7.000000e-01 & 8.000000e-01\\
      -4.913340e-01 & 6.796308e-01 & 8.000000e-01\\
      -4.646377e-01 & 4.646377e-01 & 2.000000e-01\\
      -6.796308e-01 & 4.913340e-01 & 8.000000e-01\\
      -7.000000e-01 & 1.039126e-13 & 8.000000e-01\\
      -6.796308e-01 & -4.913340e-01 & 8.000000e-01\\
      -4.646377e-01 & -4.646377e-01 & 2.000000e-01\\
      -4.913340e-01 & -6.796308e-01 & 8.000000e-01\\
      1.770206e-13 & -7.000000e-01 & 8.000000e-01\\
      4.913340e-01 & -6.796308e-01 & 8.000000e-01\\
      4.646377e-01 & -4.646377e-01 & 2.000000e-01\\
      6.796308e-01 & -4.913340e-01 & 8.000000e-01\\
      \hline
    \end{tabular}
  \end{tabular}
\end{table}

\section*{Acknowledgements}

This study was partially supported by the JSPS KAKENHI Grant number JP21H03454. In addition, the author would like to appreciate Tatsuya Iwasaki at Nagoya University for his assistance at the preliminary stage of the present study. Moreover, the author would like to thank the anonymous editor and referee for providing precious comments to the original manuscript.

\section*{Data Availability Statement}

The data that support the findings of this study are available from the corresponding author upon reasonable request.


\iftrue 

\else
\biboptions{numbers,sort&compress} 
\bibliographystyle{elsarticle/elsarticle-num}
\bibliography{reference.bib,wave3d_paper.bib} 

\begin{thebibliography}{10}
\expandafter\ifx\csname url\endcsname\relax
  \def\url#1{\texttt{#1}}\fi
\expandafter\ifx\csname urlprefix\endcsname\relax\def\urlprefix{URL }\fi
\expandafter\ifx\csname href\endcsname\relax
  \def\href#1#2{#2} \def\path#1{#1}\fi

\bibitem{turner1993}
R.~Turner,
  \href{https://www.sciencedirect.com/science/article/pii/0730725X9390209V}{Gradient
  coil design: A review of methods}, Magnetic Resonance Imaging 11~(7) (1993)
  903--920.
\newblock \href {https://doi.org/https://doi.org/10.1016/0730-725X(93)90209-V}
  {\path{doi:https://doi.org/10.1016/0730-725X(93)90209-V}}.
\newline\urlprefix\url{https://www.sciencedirect.com/science/article/pii/0730725X9390209V}

\bibitem{hidalgo2010}
S.~Hidalgo-Tobon,
  \href{https://onlinelibrary.wiley.com/doi/abs/10.1002/cmr.a.20163}{Theory of
  gradient coil design methods for magnetic resonance imaging}, Concepts in
  Magnetic Resonance Part A 36A~(4) (2010) 223--242.
\newblock \href
  {http://arxiv.org/abs/https://onlinelibrary.wiley.com/doi/pdf/10.1002/cmr.a.20163}
  {\path{arXiv:https://onlinelibrary.wiley.com/doi/pdf/10.1002/cmr.a.20163}},
  \href {https://doi.org/https://doi.org/10.1002/cmr.a.20163}
  {\path{doi:https://doi.org/10.1002/cmr.a.20163}}.
\newline\urlprefix\url{https://onlinelibrary.wiley.com/doi/abs/10.1002/cmr.a.20163}

\bibitem{smith2016}
E.~C. Smith, Advanced modelling and optimization of gradient coils and their
  physical behaviour in traditional and paired mri systems, Ph.D. thesis, The
  University of Queensland, Australia (2016).

\bibitem{turner1986target_field_approach}
R.~Turner, A target field approach to optimal coil design. j phys d, appl phys
  19:l147-l151, Journal of Physics D Applied Physics 19 (1986) L147.
\newblock \href {https://doi.org/10.1088/0022-3727/19/8/001}
  {\path{doi:10.1088/0022-3727/19/8/001}}.

\bibitem{shen2022gradient_coil_design}
S.~Shen, N.~Koonjoo, X.~Kong, M.~S. Rosen, Z.~Xu, Gradient coil design and
  optimization for an ultra-low-field {MRI} system, Applied Magnetic Resonance
  53 (2022) 895--914.
\newblock \href {https://doi.org/https://doi.org/10.1007/s00723-022-01470-2}
  {\path{doi:https://doi.org/10.1007/s00723-022-01470-2}}.

\bibitem{ryu2006IEEE}
J.~S. Ryu, Y.~Yao, C.~S. Koh, Y.~J. Shin, 3-d optimal shape design of pole
  piece in permanent magnet mri using parameterized nonlinear design
  sensitivity analysis, IEEE Transactions on Magnetics 42~(4) (2006)
  1351--1354.
\newblock \href {https://doi.org/10.1109/TMAG.2006.871563}
  {\path{doi:10.1109/TMAG.2006.871563}}.

\bibitem{sanchez2011design}
C.~Cobos~Sanchez, M.~Fernandez~Pantoja, R.~Gomez~Martin, Design of gradient
  coil for magnetic resonance imaging applying particle-swarm optimization,
  IEEE Transactions on Magnetics 47~(12) (2011) 4761--4768.
\newblock \href {https://doi.org/10.1109/TMAG.2011.2159510}
  {\path{doi:10.1109/TMAG.2011.2159510}}.

\bibitem{fernow2022principles}
R.~C. Fernow, {Principles of Magnetostatics}, Oxford University Press, 2022.
\newblock \href {https://doi.org/10.1017/9781009291156}
  {\path{doi:10.1017/9781009291156}}.

\bibitem{golay1958}
M.~J.~E. Golay, \href{https://doi.org/10.1063/1.1716184}{{Field Homogenizing
  Coils for Nuclear Spin Resonance Instrumentation}}, Review of Scientific
  Instruments 29~(4) (1958) 313--315.
\newblock \href
  {http://arxiv.org/abs/https://pubs.aip.org/aip/rsi/article-pdf/29/4/313/19027538/313\_1\_online.pdf}
  {\path{arXiv:https://pubs.aip.org/aip/rsi/article-pdf/29/4/313/19027538/313\_1\_online.pdf}},
  \href {https://doi.org/10.1063/1.1716184} {\path{doi:10.1063/1.1716184}}.
\newline\urlprefix\url{https://doi.org/10.1063/1.1716184}

\bibitem{romeo1984}
F.~Rom{\'e}o, D.~I. Hoult,
  \href{https://api.semanticscholar.org/CorpusID:24715575}{Magnet field
  profiling: Analysis and correcting coil design}, Magnetic Resonance in
  Medicine 1 (1984).
\newline\urlprefix\url{https://api.semanticscholar.org/CorpusID:24715575}

\bibitem{wong1991}
E.~C. Wong, A.~Jesmanowicz, J.~S. Hyde,
  \href{https://onlinelibrary.wiley.com/doi/abs/10.1002/mrm.1910210107}{Coil
  optimization for mri by conjugate gradient descent}, Magnetic Resonance in
  Medicine 21~(1) (1991) 39--48.
\newblock \href
  {http://arxiv.org/abs/https://onlinelibrary.wiley.com/doi/pdf/10.1002/mrm.1910210107}
  {\path{arXiv:https://onlinelibrary.wiley.com/doi/pdf/10.1002/mrm.1910210107}},
  \href {https://doi.org/https://doi.org/10.1002/mrm.1910210107}
  {\path{doi:https://doi.org/10.1002/mrm.1910210107}}.
\newline\urlprefix\url{https://onlinelibrary.wiley.com/doi/abs/10.1002/mrm.1910210107}

\bibitem{du1997studies}
Y.~P. Du, D.~L. Parker,
  \href{https://www.sciencedirect.com/science/article/pii/S0730725X9600272X}{Studies
  on the performance of circular and elliptical z-gradient coils using a
  simulated annealing algorithm}, Magnetic Resonance Imaging 15~(2) (1997)
  255--262.
\newblock \href {https://doi.org/https://doi.org/10.1016/S0730-725X(96)00272-X}
  {\path{doi:https://doi.org/10.1016/S0730-725X(96)00272-X}}.
\newline\urlprefix\url{https://www.sciencedirect.com/science/article/pii/S0730725X9600272X}

\bibitem{fisher1997design}
B.~Fisher, N.~Dillon, T.~Carpenter, L.~Hall,
  \href{https://www.sciencedirect.com/science/article/pii/S0730725X96003712}{Design
  of a biplanar gradient coil using a genetic algorithm}, Magnetic Resonance
  Imaging 15~(3) (1997) 369--376.
\newblock \href {https://doi.org/https://doi.org/10.1016/S0730-725X(96)00371-2}
  {\path{doi:https://doi.org/10.1016/S0730-725X(96)00371-2}}.
\newline\urlprefix\url{https://www.sciencedirect.com/science/article/pii/S0730725X96003712}

\bibitem{juchem2010}
\href{https://www.sciencedirect.com/science/article/pii/S1090780710000789}{Magnetic
  field modeling with a set of individual localized coils}, Journal of Magnetic
  Resonance 204~(2) (2010) 281--289.
\newblock \href {https://doi.org/https://doi.org/10.1016/j.jmr.2010.03.008}
  {\path{doi:https://doi.org/10.1016/j.jmr.2010.03.008}}.
\newline\urlprefix\url{https://www.sciencedirect.com/science/article/pii/S1090780710000789}

\bibitem{du2012designCT}
X.~Du, Z.~Zhu, L.~Zhao, G.~Zhang, F.~Ning, Z.~Liu, Z.~Hou, M.~Wang, W.~Ma,
  W.~Yao, Design of cylindrical transverse gradient coil for 1.5 t mri system,
  IEEE Transactions on Applied Superconductivity 22~(3) (2012)
  4402004--4402004.
\newblock \href {https://doi.org/10.1109/TASC.2011.2174954}
  {\path{doi:10.1109/TASC.2011.2174954}}.

\bibitem{zhang2018optimisation}
Y.~Zhang, L.~Wang, Y.~Guo, Y.~Zhang,
  \href{https://ietresearch.onlinelibrary.wiley.com/doi/abs/10.1049/iet-pel.2018.6202}{Optimisation
  of planar rectangular coil achieving uniform magnetic field distribution for
  ev wireless charging based on genetic algorithm}, IET Power Electronics
  12~(10) (2019) 2706--2712.
\newblock \href
  {http://arxiv.org/abs/https://ietresearch.onlinelibrary.wiley.com/doi/pdf/10.1049/iet-pel.2018.6202}
  {\path{arXiv:https://ietresearch.onlinelibrary.wiley.com/doi/pdf/10.1049/iet-pel.2018.6202}},
  \href {https://doi.org/https://doi.org/10.1049/iet-pel.2018.6202}
  {\path{doi:https://doi.org/10.1049/iet-pel.2018.6202}}.
\newline\urlprefix\url{https://ietresearch.onlinelibrary.wiley.com/doi/abs/10.1049/iet-pel.2018.6202}

\bibitem{zhang2018spiral}
P.~Zhang, Y.~Shi, W.~Wang, Y.~Wang, A spiral, bi-planar gradient coil design
  for open magnetic resonance imaging, Technology and Health Care 26 (2018).
\newblock \href {https://doi.org/10.3233/THC-171081}
  {\path{doi:10.3233/THC-171081}}.

\bibitem{xuan2021}
L.~Xuan, X.~Kong, J.~Wu, Y.~He, Z.~Xu, Smoothly-connected crescent transverse
  gradient coil design for {50mT MRI} system, Applied Magnetic Resonance 52
  (2021) 649--660.
\newblock \href {https://doi.org/10.1007/s00723-021-01330-5}
  {\path{doi:10.1007/s00723-021-01330-5}}.

\bibitem{brey1996}
W.~W. Brey, T.~H. Mareci, J.~Dougherty, A field-gradient coil using concentric
  return paths, Journal of magnetic resonance. Series B 112~(2) (1996)
  124--130.
\newblock \href {https://doi.org/https://doi.org/10.1006/jmrb.1996.0122}
  {\path{doi:https://doi.org/10.1006/jmrb.1996.0122}}.

\bibitem{lu2004momentum-weighted}
H.~Lu, A.~Jesmanowicz, S.-J. Li, J.~S. Hyde,
  \href{https://onlinelibrary.wiley.com/doi/abs/10.1002/mrm.10662}{Momentum-weighted
  conjugate gradient descent algorithm for gradient coil optimization},
  Magnetic Resonance in Medicine 51~(1) (2004) 158--164.
\newblock \href
  {http://arxiv.org/abs/https://onlinelibrary.wiley.com/doi/pdf/10.1002/mrm.10662}
  {\path{arXiv:https://onlinelibrary.wiley.com/doi/pdf/10.1002/mrm.10662}},
  \href {https://doi.org/https://doi.org/10.1002/mrm.10662}
  {\path{doi:https://doi.org/10.1002/mrm.10662}}.
\newline\urlprefix\url{https://onlinelibrary.wiley.com/doi/abs/10.1002/mrm.10662}

\bibitem{hughes2005}
T.~Hughes, J.~Cottrell, Y.~Bazilevs,
  \href{http://www.sciencedirect.com/science/article/pii/S0045782504005171}{Isogeometric
  analysis: {CAD}, finite elements, {NURBS}, exact geometry and mesh
  refinement}, Computer Methods in Applied Mechanics and Engineering
  194~(39--41) (2005) 4135--4195.
\newblock \href {https://doi.org/10.1016/j.cma.2004.10.008}
  {\path{doi:10.1016/j.cma.2004.10.008}}.
\newline\urlprefix\url{http://www.sciencedirect.com/science/article/pii/S0045782504005171}

\bibitem{sokolowski1991}
J.~Sokolowski, J.~Zol{\'e}sio, Introduction to shape optimization: Shape
  Sensitivity Analysis, Springer--Verlag, United States of America, 1992.

\bibitem{feynman1965flp}
R.~Feynman, R.~Leighton, M.~Sands, The Feynman Lectures on Physics:
  Electromagnetism and matter, The Feynman Lectures on Physics, Addison-Wesley
  Publishing Company, 1963.

\bibitem{hoschek1993}
J.~Hoschek, D.~Lasser,
  \href{http://www.zentralblatt-math.org/zmath/en/search/?an=0788.68002 /
  http://www.gbv.de/dms/hbz/toc/ht005011809.PDF}{Fundamentals of computer-aided
  geometric design}, Peters, Wellesley, Mass, 1993.
\newline\urlprefix\url{http://www.zentralblatt-math.org/zmath/en/search/?an=0788.68002
  / http://www.gbv.de/dms/hbz/toc/ht005011809.PDF}

\bibitem{abramowitz1972}
M.~Abramowitz, I.~Stegun (Eds.), Handbook of Mathematical Functions: with
  Formulas, Graphs, and Mathematical Tables (eighth Dover printing), Dover, New
  York, 1972.

\bibitem{takahashi2019ewco}
T.~Takahashi, T.~Yamamoto, Y.~Shimba, H.~Isakari, T.~Matsumoto,
  \href{https://doi.org/10.1007/s00366-018-0606-6}{A framework of shape
  optimisation based on the isogeometric boundary element method toward
  designing thin-silicon photovoltaic devices}, Eng. with Comput. 35~(2) (2019)
  423--449.
\newblock \href {https://doi.org/10.1007/s00366-018-0606-6}
  {\path{doi:10.1007/s00366-018-0606-6}}.
\newline\urlprefix\url{https://doi.org/10.1007/s00366-018-0606-6}

\bibitem{takahashi2023IJNME}
T.~Takahashi, N.~Miyazawa, M.~Tanigawa,
  \href{https://onlinelibrary.wiley.com/doi/abs/10.1002/nme.7130}{A
  three-dimensional shape optimization for transient acoustic scattering
  problems using the time-domain boundary element method}, International
  Journal for Numerical Methods in Engineering 124~(2) (2023) 482--512.
\newblock \href
  {http://arxiv.org/abs/https://onlinelibrary.wiley.com/doi/pdf/10.1002/nme.7130}
  {\path{arXiv:https://onlinelibrary.wiley.com/doi/pdf/10.1002/nme.7130}},
  \href {https://doi.org/https://doi.org/10.1002/nme.7130}
  {\path{doi:https://doi.org/10.1002/nme.7130}}.
\newline\urlprefix\url{https://onlinelibrary.wiley.com/doi/abs/10.1002/nme.7130}

\bibitem{takahashi2024ewco_pre}
T.~Takahashi, \href{https://doi.org/10.21203/rs.3.rs-3865606/v1}{An
  electromagnetic shape optimisation for perfectly electric conductors by the
  time-domain boundary integral equations}, Research Square (January 2024).
\newline\urlprefix\url{https://doi.org/10.21203/rs.3.rs-3865606/v1}

\bibitem{kraft1988software}
D.~Kraft, A Software Package for Sequential Quadratic Programming, Deutsche
  Forschungs- und Versuchsanstalt f{\"u}r Luft- und Raumfahrt K{\"o}ln:
  Forschungsbericht, Wiss. Berichtswesen d. DFVLR, 1988.

\bibitem{nlopt}
S.~G. Johnson, \href{http://github.com/stevengj/nlopt}{The {NLopt}
  nonlinear-optimization package}, http://github.com/stevengj/nlopt.
\newline\urlprefix\url{http://github.com/stevengj/nlopt}

\bibitem{powell2007}
M.~J.~D. Powell,
  \href{http://www.damtp.cam.ac.uk/user/na/NA_papers/NA2007_03.pdf}{A view of
  algorithms for optimization without derivatives}, Cambridge Uinversity
  Technical Report (2007) 10--12.
\newline\urlprefix\url{http://www.damtp.cam.ac.uk/user/na/NA_papers/NA2007_03.pdf}

\bibitem{palmer2006new_fMRI}
A.~R. Palmer, J.~Chambers, D.~A. Hall, New fmri methods for hearing and speech,
  Acoustical Science and Technology 27~(3) (2006) 125--133.
\newblock \href {https://doi.org/10.1250/ast.27.125}
  {\path{doi:10.1250/ast.27.125}}.

\bibitem{motovilova2022}
E.~Motovilova, S.~A. Winkler,
  \href{https://www.frontiersin.org/articles/10.3389/fphy.2022.907619}{Overview
  of methods for noise and heat reduction in mri gradient coils}, Frontiers in
  Physics 10 (2022).
\newblock \href {https://doi.org/10.3389/fphy.2022.907619}
  {\path{doi:10.3389/fphy.2022.907619}}.
\newline\urlprefix\url{https://www.frontiersin.org/articles/10.3389/fphy.2022.907619}

\bibitem{mansfield1995}
P.~Mansfield, B.~L. Chapman, R.~Bowtell, P.~Glover, R.~Coxon, P.~R. Harvey,
  Active acoustic screening: reduction of noise in gradient coils by {Lorentz}
  force balancing, Magnetic resonance in medicine 33~(2)  276--281.
\newblock \href {https://doi.org/https://doi.org/10.1002/mrm.1910330220}
  {\path{doi:https://doi.org/10.1002/mrm.1910330220}}.

\end{thebibliography}
\fi

\end{document}